\documentclass[a4paper,usenatbib]{mn2e}
\usepackage{hyperref,graphicx}
\usepackage{amsmath,amssymb}

\hypersetup{colorlinks=true, citecolor=blue, linkcolor= magenta}

\newcommand{\apj}{ApJ}
\newcommand{\apjl}{ApJ}

\newcommand{\mnras}{MNRAS}
\newcommand{\nat}{Nature}

\newcommand{\aap}{A{\&}A}

\newcommand{\memsai}{Memorie della Societa Astronomica Italiana}

\newcommand{\actaa}{Acta Astronomica}
\newcommand{\nustar}{NuSTAR J09551$+$6940.8}
\newcommand{\prd}{Phys. Rev. D.}

\title[The ULX NuSTAR J095551+6940.8: A magnetar in a HMXB]
{The ultra luminous X-ray source NuSTAR J095551+6940.8: A magnetar in a high mass X-ray binary}

\author[Ek\c{s}i et al.\ ]
{K.\ Y.\ Ek\c{s}i, \.{I}.\ C.\ Anda\c{c}, S.\ \c{C}{\i}k{\i}nto\u{g}lu, A.\ A.\ Gen\c{c}ali, C.\ G\"{u}ng\"{o}r \& F.\ \"{O}ztekin \\  
  Istanbul Technical University,
  Faculty  of Science  and  Letters,  Physics Engineering  Department,
  34469,  Istanbul, Turkey}  

\begin{document}
\date{please contact: \url{eksi@itu.edu.tr}}
\pagerange{\pageref{firstpage}--\pageref{lastpage}} \pubyear{2014}
\maketitle

\begin{abstract}

The recent detection of pulsations from the ultra luminous X-ray source (ULX)
NuSTAR J095551+6940.8 in M82 by Bachetti et al.\ indicates that the 
object is an accreting neutron star in a high mass X-ray binary (HMXB) system. 
The super-Eddington luminosity of the object implies that the magnetic field is 
sufficiently strong to suppress the scattering cross-section unless its beam is viewed at a favourable angle. 
We show that the torque equilibrium condition for the pulsar indicates the dipole magnetic field of the neutron star is $6.7 \times 10^{13}$~G, two orders of magnitude higher than that estimated by Bachetti et al.,  
and further point to the possibility that even
stronger magnetic fields could well be in the higher multipoles. This supports the recent view that magnetars descent from HMXBs if the magnetic field decays an order of magnitude during the process of transition.
\end{abstract}


\label{firstpage}

\section{INTRODUCTION}
\label{intro}

Ultra-luminous X-ray sources (ULXs) are accreting compact objects with luminosities exceeding the Eddington 
limit $L_{\rm E} = 4\pi G M m_p c/\sigma_{\rm T} =1.3\times 10^{38}(M/M_{\odot})$~erg~s$^{-1}$ 
($M$ is the mass of the accreting object, $m_p$ is the mass of the proton, $c$ is the speed of light, 
$G$ is the gravitational constant and $\sigma_{\rm T}$ is the Thomson cross-section for scattering of photons from electrons) 
for a $\sim 10M_{\odot}$ object \citep[see][for a review]{rob07}. 
They possibly form a heterogeneous family with sub-classes \citep{gla13}.  
Some of these objects reach luminosities $10^{41}$~erg~s$^{-1}$ appropriate 
for $M \sim 10^3~M_{\odot}$ under the assumption of accretion at the Eddington limit. 
Such objects are puzzling as they can not form by stellar evolution and are dubbed 
intermediate mass black holes \citep{kon+04,mil+04,liu08}. 
Alternatively, they are proposed to be stellar mass black holes accreting at rates 
slightly exceeding the Eddington limit \citep{gla+09} and appear to be super-Eddington due to anisotropic emission. 

The recent identification \citep{bach+14} of  \nustar~  in the nearby galaxy M82 
with a neutron star accreting in a high mass X-ray binary (HMXB) challenged the view 
that all ULX harbour black holes. 
The luminosity of the object is $4\times 10^{39}$~erg~s$^{-1}$ during pulsations 
and reaches $3.7\times 10^{40}$~erg~s$^{-1}$ at the peak flux under the assumption of isotropic emission. 
As a neutron star can not have a mass exceeding $\sim 3.5M_{\odot}$ \citep{rho74} 
this super-Eddington luminosity is attributed by \citet{bach+14} to fan beam geometry \citep{gne73} 
viewed at a favourable angle. An alternative explanation obviating the assumption of favourable viewing angle though not necessarily rejecting that the emission is anisotropic, is that the magnetic field 
is so strong that the scattering cross-section reduces \citep{can71} as is 
the case for magnetars \citep{dun92,tho96,pac92}. This explanation is not preferred by \citep{bach+14} 
as they infer the magnetic dipole field of the neutron star to be $B \sim 10^{12}$~G from 
the observed spin-up rate assuming accretion at the Eddington rate. 
This value for the magnetic field inferred by \citep{bach+14} is not consistent with the assumption of the system being near torque equilibrium
because the torque for a system near equilibrium is vanishingly small and leads to an underestimate
of the magnetic field if its full value is employed. A magnetic field as low as $B \sim 10^{12}$~G  is also not consistent with the presence of fan beam geometry. An increase in the critical luminosity by beaming is easier to achieve with a field of the order of $10^{13}$~G \citep{bask75}.

In this \textit{letter} we show that  the pulsar in \nustar~ has a dipole magnetic field of at least $B\sim 10^{13}$~G---in the range of low-magnetic field magnetars \citep{rea+10}---and possibly
of $B \sim 10^{14}$~G in excess of the quantum critical limit,
$B_{\rm c} \equiv m_e^2 c^3/\hbar e= 4.4\times 10^{13}$~G, and argue that it could have even stronger 
magnetic fields in the higher multipoles \citep{eks03}. 
Such strong fields increase the
critical luminosity either by releasing the energy via magnetohydrodynamic waves \citep{kat96} or by the reduction of the scattering cross section  \citep{can71}. 
This relates at least some of the ULXs with isolated magnetars and supports the recent view that magnetars descent from HMXBs \citep{bis14}. In the following section we estimate the magnetic field of the accreting neutron star in \nustar~ and in \S 3 we discuss its astrophysical implications.

\section{The magnetic field of \nustar~ and its critical luminosity}
\label{mag}

The X-ray luminosity due to accretion of matter onto a compact object is
\begin{equation}
L_{\rm X} = \frac{GM\dot{M}}{R}
\end{equation}
where $R$ is the radius of the neutron star. This implies $\dot{M}=0.535\times 10^{20}$~g~s$^{-1}L_{40}R_6 m^{-1}$ 
where $L_{40}=L/10^{40}$~erg~s$^{-1}$, $R_6 =R/10^6$~cm and $m=M/1.4M_{\odot}$. 
Normally, this much luminosity would not be able to accrete onto the star because 
of the radiation pressure. Yet, we assume the critical luminosity is 
increased beyond the Eddington limit by the suppressed scattering cross-section of the electrons in strong magnetic field \citep{can71,her79,pac92} or by the transportation of energy via magnetohydrodynamic waves \citep{kat96}.

\begin{figure}
\centering
  \includegraphics[width=0.5\textwidth]{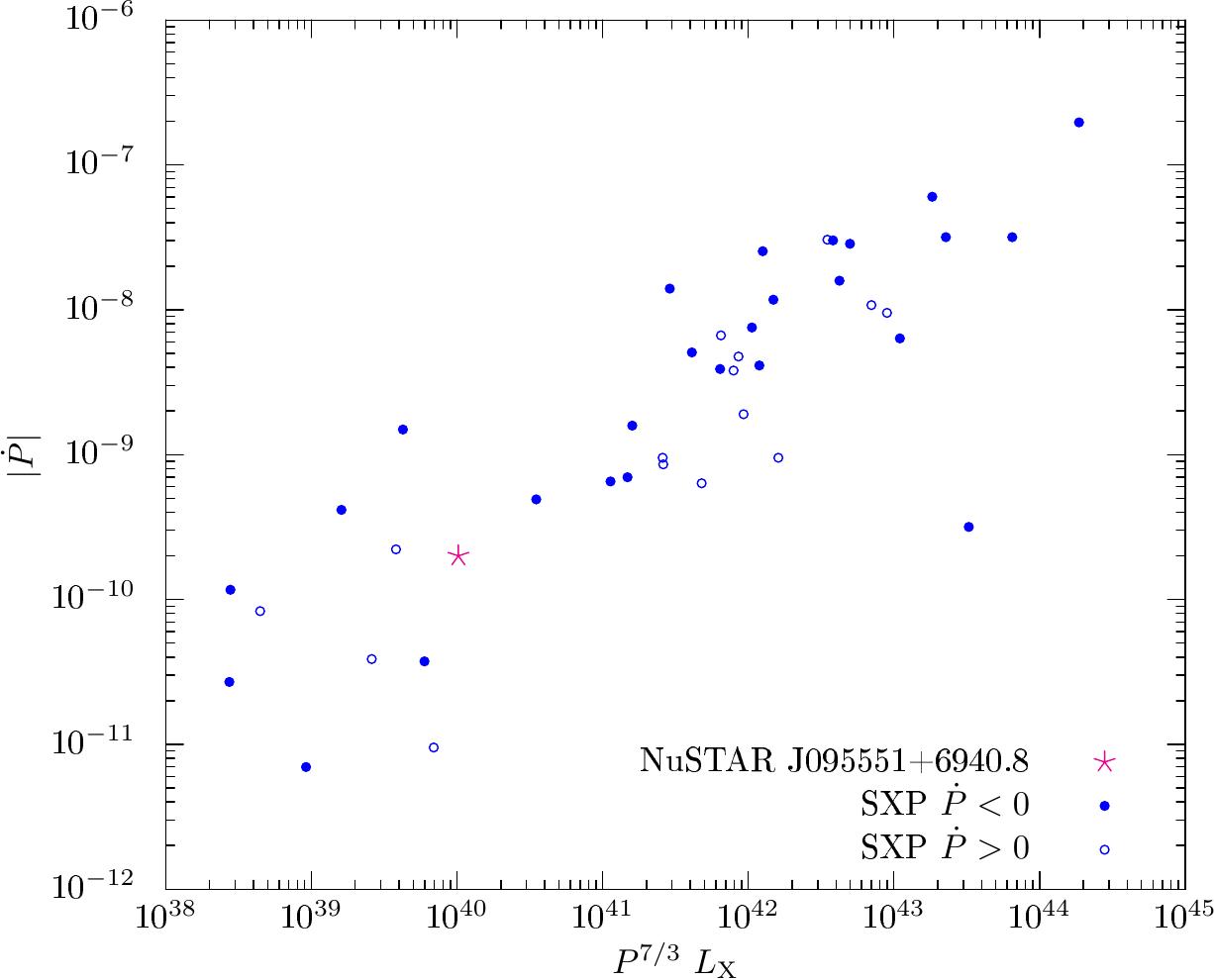}
     \caption{Period derivative $|\dot{P}|$ versus $L_{\rm X}P^{7/3}$ diagram \citet{gho79b} 
     for X-ray pulsars in Be/HMXBs of Small Magellanic Clouds \citep{klu+14} (SXPs) together with \nustar~ where we used $P=1.37$, $\dot{P}=2\times 10^{-10}$~s~s$^{-1}$ and $L_{\rm X}= 4.9\times 10^{39}$~erg~s$^{-1}$ \citep{bach+14}.}
         \label{fig1}
\end{figure}

The torque on the neutron star is $N = I \dot{\Omega}=6\times 10^{35}I_{45}$~g~cm$^2$~s$^{-2}$ 
where $I_{45}$ is the moment of inertia in units of 10$^{45}$~g~cm$^2$ as inferred from 
the observed spin-up rate of $\dot{P} \sim -2 \times 10^{-10}$~s~s$^{-1}$ \citep{bach+14}. 
As $P/\dot{P}\sim 300$~years the system should be near torque equilibrium so that the inner radius of the disc is
close to the corotation radius $R_{\rm c} = (GMP^2/4\pi^2)^{1/3}$. 
More specifically it will be $R_{\rm in}=\omega_{\rm c}^{2/3}R_{\rm c}$ where $\omega_{\rm c}$ 
is the critical fastness parameter at which torque vanishes and is of order unity. 
The fastness parameter is defined as $\omega_{\rm \ast}=(R_{\rm in}/R_{\rm c})^{3/2}$ \citep{els77}. 
The measured period of 
the neutron star, $P=1.37$~s, implies that the co-rotation radius is  at
$R_{\rm c}=2.1\times 10^8 m^{3/2}~{\rm cm}$. The inner radius of 
the disc is thus at
\begin{equation}
R_{\rm in} = 2.1\times 10^8 m^{3/2}\omega_{\rm c}^{2/3}~{\rm cm}.
\end{equation}
The inner radius of the disc is determined by the balance of magnetic and material stresses \citep{gho79a,gho79b} and scales with the Alfv\'en radius \citep{dav73}
\begin{equation}
R_{\rm in} = \xi \left( \frac{\mu^2}{\sqrt{2GM}\dot{M}} \right)^{2/7}
\end{equation}
where $\xi$ is a dimensionless number of order unity. 
Equating the last two equations we find the magnetic dipole moment of the neutron star as 
$\mu=1.17\times 10^{31}\omega_{\rm c}^{7/6} \xi^{-7/4} m^{1/3}L_{40}^{1/2}R_6^{1/2}$ 
which, by $\mu=\frac12 BR^3$, implies 
\begin{equation}
B= 2 \times 10^{13}\omega_{\rm c}^{7/6} \xi^{-7/4} m^{1/3}L_{40}^{1/2}R_6^{-5/2}~{\rm G}.
\label{Bfield}
\end{equation}
This value is an order of magnitude larger than
that found by  \citet{bach+14} who use the measured torque to estimate the magnetic field. 
For $\xi=0.5$, favoured by \citet{gho79a}, the field is found to be even larger, $B = 6.7\times 10^{13}$~G,  in  excess of 
the quantum critical limit $B_{\rm c}$. 
The torque, however, is smaller than its nominal value $N_0 = \sqrt{GMR_{\rm in}}\dot{M}$ \citep{pri72}
as the system is near torque equilibrium. The nominal value should not be used for estimating the magnetic moment in this case as it would give a lowest estimate for the magnetic field.
The torque acted by the disc, in general, is given as
\begin{equation}
N= n(\omega_{\rm \ast})N_0
\end{equation}
where $n(\omega_{\rm \ast})$ is the dimensionless torque \citep{gho79a,gho79b}. Any torque model near torque equilibrium would be of the form $n=1-\omega_{\ast}/\omega_{\rm c}$. This then would lead to the equation
\begin{equation}
-I\dot{P}/P^2 = \left(1-\frac{\omega_{\ast}}{\omega_{\rm c}} \right) \omega_{\ast}^{1/3} \sqrt{GMR_{\rm c}}\dot{M}
\end{equation} 
where we used $N=-I\dot{P}/P^2$. We solve this equation which is of the form $(1-x)x^{1/3}={\rm constant}$ where $x=\omega_{\ast}/\omega_{\rm c}$ numerically for the fastness parameter of the system and find $\omega_{\ast} \simeq 0.9$ for $\omega_{\rm c}=1$, $P=1.37$, $\dot{P}=2\times 10^{-10}$~s~s$^{-1}$ and $L_{\rm X}= 4.9\times 10^{39}$~erg~s$^{-1}$ \citep{bach+14}. This indeed shows that the system is near torque equilibrium, but leads to a magnetic field $0.9^{7/6}\simeq 0.88$ times smaller than that inferred in \autoref{Bfield}.
The above relations lead to
\begin{equation}
|\dot{P}| \propto (L_{\rm X}P^{7/3})^{6/7}
\end{equation} 
\citep{gho79b} for $\omega_{\ast} \ll \omega_{\rm c}$ and this relation may be used as a diagnostic for understanding the accretion stage of \nustar.
Accordingly, Figure~\ref{fig1} shows \nustar~ among X-ray pulsars in Be/HMXBs of the 
Small Magellanic Clouds  (SXPs) \citep{klu+14} in a $|\dot{P}|$ versus $L_{\rm X}P^{7/3}$ digram. The SXPs in the data set of \citet{klu+14} are the ones accreting from a disc rather than a wind.
The ordinary position of \nustar~ in the diagram also suggests that the object is indeed near torque
equilibrium just like most of the SXPs though \nustar is likely to accrete via Roche-lobe overflow while SXP would be accreting from Be discs.

\section{Discussion and Conclusion}
\label{disc}

We have found that the dipole magnetic field of the pulsar \nustar~ is at least $2 \times 10^{13}$~G, an order of magnitude larger than that inferred by \citep{bach+14} and possibly even stronger, $6.7 \times 10^{13}$~G,  in excess of 
the quantum critical magnetic field $B_{\rm c}$, if the inner radius of the disc is half of the Alfv\'en radius. Such a super-strong field is 
sufficient to lead to a reduction of the scattering 
cross-section and increase the critical luminosity \citep{pac92}. It is possible that the object has 
even stronger magnetic fields in the higher multipoles leading to even more effective 
increase of the critical luminosity enabling luminosities as large as $10^3 L_{\rm E}$.

Such strong fields are reminiscent of the magnetars \citep{dun92} in our own galaxy which are
envisaged as isolated objects \citep{tho96}. Recently, \citet{bis14} suggested
a scenario in which magnetars descent from HMXBs. The magnetic fields suggested for magnetars by these authors, however, is an order of magnitude lower than that estimated for \nustar~ in the present paper. This implies that, if the scenario is correct, the magnetic field of \nustar~ should decay during the process of transition to the isolated magnetar stage. 
The decay of the magnetic field of a neutron star by accretion is a process also invoked in the context of millisecond pulsars descending from low mass X-ray binaries \citep{bis76,alp+82,rad82}. 

There had been claims for the presence of accreting neutron stars with  such strong magnetic fields in other HMXBs \citep[see e.g.]{rei+12,boz+08,pop12,fin+10,ikh10,klu+13,fu12} though none had been caught in such a ultra-luminous state as \nustar.
Also the neutron stars in these systems are accreting from the wind of the companion and have very long spin periods, $P \sim 10^3$~s.
More interestingly, magnetar-like
behaviour from the peculiar binary LS~I~$+$61{$^\circ$}303 is reported \citep{tor+12}. 
The presence of such strong magnetic fields in some of these systems, however, has recently been questioned by \citep{pos+14} who employed the quasi-spherical accretion model of \citet{sha+12}. \nustar, however, is very different from these systems with its much shorter period and much higher luminosity.
 
For such strong magnetic fields the electron-cyclotron line would be beyond 500 keV which is undetectable by X-ray detectors. The proton-cyclotron lines have energies $E \sim 0.5(B/10^{14}~{\rm G})=0.5$~keV and could be detected by \textit{XMM-Newton}. We would, however, like to note that no significant proton-cyclotron line had been detected in the persistent X-ray emission of isolated magnetars. Hence the null result of such a search would not immediately rule out that the magnetic field of this object is of magnetar strength.

The question naturally arises whether most if not all ULXs could be super-strongly magnetized accreting neutron stars. Soon after this paper was submitted \citet{dor+14} presented the null result of their search of pulsations through available  archival \textit{XMM-Newton} observations of several ULX.
At highest accretion rates the pulsar could be enshrouded and the pulsations could be 
smeared out by the optically thick surrounding medium. This may address the elusiveness  
of pulsations from other ULXs and may indicate to a larger population of neutron stars 
among ULXs. Accordingly, ULXs at their lowest luminosity stage are better targets for 
detecting pulsations.
 

\section*{Acknowledgements}
KYE, CG, CA and S{\c C} acknowledge support from The Scientific and Technological Council of TURKEY 
(TUBITAK) with the project number 112T105.

\bibliographystyle{mn2e}


\label{lastpage}

\end{document}